\documentclass[sort&compress,1p]{elsarticle}
\usepackage{amsmath,amssymb,bm}
\usepackage{graphicx}\graphicspath{{Graph/}{../Graph/}{../../Graph/}{}}
\usepackage[T1]{fontenc}
\usepackage[utf8]{inputenc}
\usepackage[english,american]{babel}

\usepackage[
    colorlinks
        ,linkcolor=violet
        ,filecolor=purple
        ,citecolor=teal
    ,unicode
    ,bookmarksnumbered
    ,pdfpagemode=UseOutlines% (show bookmarks)
    ,pdfdisplaydoctitle=true% display document title instead of file
    ,bookmarksopenlevel = 1
]{hyperref}
    \hypersetup{
        pdftitle={Electromagnetic surface waves on a conducting cylinder}
        ,pdfauthor={Igor A. Kotelnikov, Gennady V. Stupakov},
        ,pdfsubject={Draft, \today}
        ,pdfkeywords={polariton-plasmon, surface wave}
    }

\DeclareMathOperator{\re}{Re}
\DeclareMathOperator{\im}{Im}

\DeclareMathOperator{\e}{e}

\DeclareMathOperator{\rot}{rot}

\DeclareMathOperator{\Ai}{Ai} \DeclareMathOperator{\Bi}{Bi}

\DeclareMathOperator{\Hankel}{H}

\providecommand{\bm}[1]{\mbox{\boldmath$#1$}}
\renewcommand{\vec}[1]{\bm{#1}}
\renewcommand{\leq}{\leqslant}

\newcommand{\parder}[2]{\frac{\partial #1}{\partial #2}}
\newcommand{\tparder}[2]{{\partial #1}/{\partial #2}}
\newcommand{\dif}[1][]{\mathop{}\!\mathrm{d}
    \if\relax\detokenize{#1}\relax\else
        ^{\mkern-1.mu#1}\mkern-2.5mu
    \fi}
\newcommand{\der}[2]{\frac{\dif{#1}}{\dif{#2}}}

    \usepackage{xcolor}
    \definecolor{darkblue}{cmyk}{1.00, 0.50, 0.00, 0.40}

\begin{document}

%\thispagestyle{empty}
%\renewcommand{\thefootnote}{\fnsymbol{footnote}}

%Title of paper
%% Version:
%% Dispersion of the surface modes on a curved air-metal interface
%\title{Paraxial equation for the electromagnetic surface wave}
\title{Electromagnetic surface waves on a conducting cylinder}

\author[binp,nsu]{Igor A. Kotelnikov}
\cortext[mycorrespondingauthor]{Corresponding author}
%\email{I.A.Kotelnikov@inp.nsk.su}
\address[binp]{Budker Institute of Nuclear Physics SB RAS,\\
Lavrentyev Avenue 11, Novosibirsk, 630090, Russia}

\address[nsu]{Novosibirsk State University,\\
Pirogova Street 11, Novosibirsk, 630090, Russia}

\author[slac]{Gennady V. Stupakov}
%\email{stupakov@slac.stanford.edu}
\address[slac]{SLAC National Accelerator Laboratory,\\
2575 Sand Hill Road, Menlo Park, CA, 94025, USA}

\begin{abstract}
    %We study propagation of electromagnetic surface waves on a metal-air interface in the case when the wave frequency $\omega $ is below the plasma frequency $\omega_{p}$. We derive a reduced wave equation for a metal-air boundary with a given radius of curvature $R$. Using the Leontovich boundary condition we find solutions to this equation which we classify as outgoing and incoming surface waves. We argue that the earlier studies overlooked the waves of the second type although they are the only type of surface waves which can propagate on a planar metal-air boundary. We derive the dispersion relations of the surface waves of both types and show how the dispersion relation of a planar surface wave is recovered as the limit of a sufficiently large curvature radius $R$. Our results are important for correct interpretation of experiments on generation and propagation of terahertz surface electromagnetic waves along curved metal surfaces.

    We study propagation of electromagnetic surface waves on a metal-air interface in the case when the wave frequency is below the plasma frequency. We derive a reduced wave equation for a metal cylinder with a given radius of curvature. Using the Leontovich boundary condition we find solutions to this equation which we classify as outgoing and incoming surface waves. We derive the dispersion relations of the surface waves of both types and argue that the earlier studies overlooked the waves of the second type although they are the only type which can propagate on a planar metal-air boundary.

\end{abstract}

%
%% REPLACE WITH CORRECT OCIS CODES FOR YOUR ARTICLE
%% NOTE: \ocis{} IS ALIASED TO \pacs{} BUT MUST
%% FORMAT THE TERMS CORRECTLY FOR EACH JOURNAL
%\ocis{
%    (240.6680) Surface plasmons;
%    (260.3910) Metal optics;
%    (240.6690) Surface waves.
%}
%\pacs{
%    % 42.25.-p	Wave optics
%    42.25.Gy, % Edge and boundary effects; reflection and refraction
%    42.25.Bs, % Optical absorption, wave propagation,
%    73.20.Mf, % Surface plasmons,
%    78.67.-n  % Low-dimensional structures, optical properties
%    %42.79.Gn  % Cavity resonators, optical,
%}


\maketitle

\section{Introduction}
\label{s1}

Recent developments that lead to extending of plasmonics to the mid- and far-infrared (terahertz) ranges of frequencies constitute an important direction of modern photonics \cite{Mills1982surface, Maier2007, Stanley2012NP_6_409}. Surface Plasmon Polaritons (SPPs)  are partially longitudinal $p$-polarized evanescent electromagnetic waves that propagate along a metal-air interface as a result of collective electron oscillations coupled to an external electromagnetic field \cite{Zhizhin+1975UFN_18_927, Zhizhin+1982, Raether1988, Zayats+2005PhysRep_408_131, Maier2007, Novotny2012Principles, Gerasimov+2012b, Kotelnikov+2013PRA_87_023828}. In practical applications it is important to understand how SPPs propagate along a curved surface \cite{Bogomolov+2009NIM_603_52, Gong+2009OE_17_17088, Stanley2012NP_6_409}. For this reason, the theory of SPPs on a cylinder attracted a lot of attention in the past and recent years~\cite{Berry1975JPhysA_8_1952, SchroterDereux2001PhysRevB_64_125420, Hasegawa+2004ApplPhysLett_84_1835, Hasegawa2007PhysRevA.75.063816, Liaw+2008OE.16.4945, Guasoni2011JOSAB.28.1396, Guo+2009IEEE.8.408, Polanco+OptExpress2014_120}. A dispersion equation for the transverse magnetic (TM) modes was derived by M.~Berry in an artificial case of lossless metal \cite{Berry1975JPhysA_8_1952} and later by K.~Hasegawa et al. in a general case \cite{Hasegawa+2004ApplPhysLett_84_1835, Hasegawa2007PhysRevA.75.063816}. A similar equation appears in the theory of diffraction of a wave on a transparent cylinder \cite{Rulf1967JMP_8_1785, Chen1964JMP_5_820, StreiferKodis1962solution, StreiferKodis1965solution}. It involves the Bessel and Hankel functions of a large argument and a large complex order which makes its numerical evaluation for practical problems difficult.  Different approaches have been developed to overcome this difficulty as reviewed in Ref.~\cite{Guasoni2011JOSAB.28.1396} but none of them was especially successful.

The standard approach to the derivation of SPPs~\cite{Berry1975JPhysA_8_1952, Hasegawa+2004ApplPhysLett_84_1835, Hasegawa2007PhysRevA.75.063816} assumes that the radial distribution of the magnetic field of a plasmon near the surface of a metallic cylinder is described by the Hankel function of the first kind, $\Hankel_{n}^{(1)}(k_0 r)$, with $k_0 = \omega/c$, $\omega$ the frequency of the mode, and $r$ the radius in cylindrical system of coordinates. The Hankel function of the second kind, $\Hankel_{n}^{(2)}(k_0 r)$, is explicitly or implicitly discarded.

This approach is usually justified by the fact that at large distances, $k_0 r \gg 1$, the asymptotic representation of the Hankel function $\Hankel_{n}^{(1)}$ describes a radial distribution of a wave propagating away from the cylinder, whereas $\Hankel_{n}^{(2)}(k_0 r)$ stands for a wave, propagating to the cylinder. The situations however becomes more complicated if one takes into account that, in the case of finite conductivity of the metal, the index $n$, which has to be found from a dispersion equation, is a large complex number. In this case, $\Hankel_{n}^{(1)}$ can experience a large exponential growth in the radial direction in the vicinity of the metal surface, meaning that the mode is not actually localized near the surface.

In our opinion, an adequate approach to the study of SPPs is to analyze their spatial (radial plus azimuthal) distribution in the vicinity of the metal surface, $r-R \ll R$, where $R$ is the cylinder radius (see Fig.~\ref{fig:geometry} below).  We introduce a small parameter $\delta=\sqrt{a/R}$, where  $a \sim r-R$ is
%the characteristic transverse extension
a transverse localization length
of the mode and derive equations in which small terms in parameter $\delta$ are systematically neglected. A similar method was previously successfully used in the theory of synchrotron radiation in a curved waveguide \cite{Stupakov+2003PhysRevSTAB.6.034401, Stupakov+Phys2009RevSTAB.12.104401}. Here we use it for investigation of the Surface electromagnetic Waves (SW) on a  metal-air interface with given and fixed radius of curvature. Solving these equations we obtain expressions for the fields in terms of the Airy-Fock functions, avoiding in this way the numerical difficulties arising when one deals with the Hankel functions of large argument and large order.

In this our study we find that, indeed, the localized solutions in the form of $\Hankel_{n}^{(1)}$ used in~\cite{Berry1975JPhysA_8_1952, Hasegawa+2004ApplPhysLett_84_1835, Hasegawa2007PhysRevA.75.063816} and some subsequent papers (see eg. \cite{Liaw+2008OE.16.4945, Guasoni2011JOSAB.28.1396}) may exist only in a narrow range of values of the dimensionless surface impedance $\xi$. On the contrary, we find that SWs which are expressed through the Hankel functions of the second kind $\Hankel_{n}^{(2)}(k_0r)$, do exist for any $\xi$ provided that the radius of the cylinder is not too small. We named these two types of surface waves correspondingly the outgoing and incoming surface waves. We also note that in a planar geometry an ordinary surface wave belongs to the type of incoming surface waves. As to the outgoing modes discussed in Refs.~\cite{Hasegawa+2004ApplPhysLett_84_1835, Hasegawa2007PhysRevA.75.063816, Liaw+2008OE.16.4945, Guo+2009IEEE.8.408, Guasoni2011JOSAB.28.1396}, they can be identified as creeping waves (CW). In contrast to the surface waves, the amplitude of CWs grows in radial direction in the vicinity of the metal; in Ref.~\cite{Polanco+OptExpress2014_120} these waves were characterized as wave-guide modes.

The term ``creeping waves'' is used in the theory of diffraction \cite{Honl1961Springer, Rulf1967JMP_8_1785, PaknysWang1986AP_34_674, Andronov+1995JEWA_9_503, PaknysJackson2005IEEETransAP_53_898, JosefssonPersson2006Conformal, Andronov2008AJ_54_139} for the refracted waves over convex conducting or dielectric surfaces where an extra attenuation due to the curvature is present. The foundations of the theory were laid out in earlier works of V.A. Fock \cite{Fock1946b(eng), Fock1965Pergamon} who considered propagation of radio waves around the Earth.
%
%A creeping wave is characterized by a rapidly increased attenuation when the radius is decreased.
%
%A creeping wave attains a maximal amplitude at some (maybe large) distance from the surface of the metallic cylinder.
% As we mentioned in the Introduction, the term ``creeping wave''  is used in modern theory of diffraction \cite{Honl1961Springer, Rulf1967JMP_8_1785, Rulf1967JMP_8_1785, PaknysWang1986AP_34_674, Andronov+1995JEWA_9_503, PaknysJackson2005IEEETransAP_53_898, JosefssonPersson2006Conformal, Andronov2008AJ_54_139} for a refracted wave on convex surfaces, where an extra attenuation due to the surface curvature is present.
As explained in Ref.~\cite[chapt.~5]{Honl1961Springer}, the creeping waves rapidly damp in the shadow region created by a convex obstacle on the way of a wave propagation. They damp both in the direction of propagation and towards to the surface of the obstacle. In other words, the amplitude of a creeping wave grows with the distance along the normal to the surface from the shadow region to the lit one. On the contrary, the amplitude of the surface waves decays with the distance from the surface. This qualitative difference of SWs from CWs, which are also known as ``Whatson modes'', is discussed in Ref.~\cite{PaknysJackson2005IEEETransAP_53_898}.

%Creeping waves are unique for the curved surface and cannot be found in the planar case. Surface waves are not discussed in the theory of diffraction, perhaps, because of the fact that the cylinder is usually assumed to be an ideal conductor, and, as we will see, SW does not exist in this case.

The paper is organized as follows.
%The main equations are derived in Secs.~\ref{s2} and~\ref{s3}.
A paraxial equation for electromagnetic waves near a curved surface is derived in Sec.~\ref{s2}. It is used in Sec.~\ref{s3} to obtain dispersion equations of various  surface electromagnetic waves on a curved surface; we introduce here the notion of outgoing and incoming surface waves.
%and argue that incoming surface waves were overlooked in existing literature.
A numerical solution to these equations is discussed in Sec.~\ref{s4}. Analytical solutions are presented in Sec.~\ref{s5}. All these sections deal with a convex geometry, where the metallic medium is a cylinder surrounded by air. A concave case is considered in Sec.~\ref{s6}, where it is assumed that a hollow conducting cylinder is filled by air.  In Sec.~\ref{s7} we show how the Airy-Fock functions are matched to the Hankel functions in a relevant range of parameters thus providing a bridge between our results and previous studies in Ref.~\cite{Hasegawa+2004ApplPhysLett_84_1835, Hasegawa2007PhysRevA.75.063816, Liaw+2008OE.16.4945, Guasoni2011JOSAB.28.1396, Guo+2009IEEE.8.408}. Finally, in Sec.~\ref{s8} we summarize main results.

%%%%%%%%%%%%%%%%%%%%%%%%%%%%%%%%%%%
%
%\section{Paraxial equations for electromagnetic field}
\section{Paraxial equation}
\label{s2}
%
%%%%%%%%%%%%%%%%%%%%%%%%%%%%%%%%%%%

\begin{figure}[t]
    \centering
    \includegraphics[width=0.7\columnwidth]{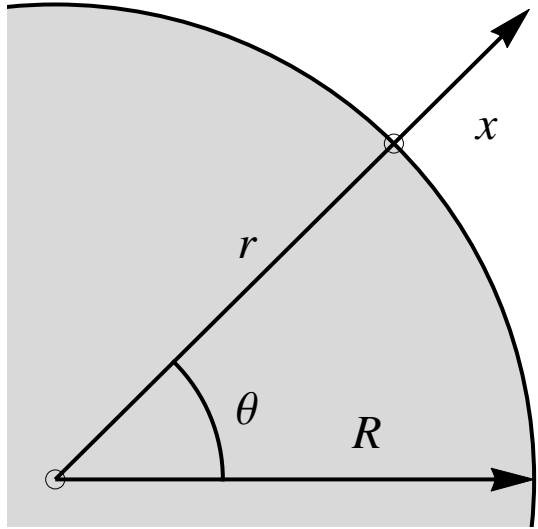}
    \caption{
        Coordinate system with $R$ the cylinder radius. A conducting medium occupies either an interior (convex geometry) or exterior (concave geometry) of the cylinder. The radial coordinate $r$ is measured from the axis of the cylinder, and $x=r-R$ in counted from the cylinder surface. A transverse magnetic (TM) wave under consideration has magnetic field $H_{z}$ directed along the axis $z$ of the cylinder.
    }
    \label{fig:geometry}
\end{figure}

In this section we will derive equations for the electric and magnetic fields near the surface of a conducting cylinder. The geometry of the problem and the choice of the coordinate system is shown in Fig.~\ref{fig:geometry}. We use the cylindrical coordinate system $r$, $\theta$, $z$ and the notation $x$ for the difference $x=r-R$. The metal surface is located at $x=0$.

We first assume that the metal occupies the region $x<0$; the region $x>0$ has the dielectric permeability $\varepsilon = 1$.
% with $\varepsilon = 1$.
This is so called \emph{convex case} \cite{Liaw+2008OE.16.4945}, where a metallic circular cylinder is surrounded by an empty space (or by air). Later on, we will extend our results to the \emph{concave case} by assigning an opposite sign to the radius $R$. As to the current and subsequent sections, except for Sec.~\ref{s6}, we suppose that $R>0$.

We assume that all components of the field have the following dependence on time and the azimuthal angle $\theta$
    \begin{gather}
    \vec{E},\vec{H}
    \propto
    \vec{U}(r)
    \e^{-i\omega t+ikR\theta}
    \end{gather}
with amplitude $\vec{U}(r)$ independent of the coordinate $z$. %{\color{red}Moreover, we restrict ourselves to the case of transverse magnetic (TM) wave since we expect that there exist no transverse electric (TE) surface waves; at least, this is the case in planar geometry. \tiny GS: this is not clear} Hence, the magnetic field
Moreover, we restrict ourselves to the case of transverse magnetic (TM) waves
%
%waves since there exist no transverse electric (TE) surface waves in the case of planar geometry (see eg. \cite{Mills1982surface, Maier2007, Raether1988, Sernelius2001, Klimov2009eng}).
%
although it has been recently reported \cite{Polanco+2012PhysLettA_1573} that transverse electric (TE) surface waves may exist on a cylindrical metal-air interface in contrast to the absence of such waves in a planar geometry (see eg. \cite{Mills1982surface, Maier2007, Raether1988, Sernelius2001, Klimov2009eng}).
Hence, we assume that the magnetic field
    \begin{gather*}
    \vec{H}=\{0,0,H_{z}\}
    \end{gather*}
has only the $z$-component, and the electric field
    \begin{gather*}
    \vec{E}=\{E_{r},E_{\theta},0\}
    \end{gather*}
has both the $r$- and $\theta$-components.

To keep track of order of magnitudes, we will assume that $x\sim a$ is of the order of $\delta^{0}$ and the radius $R$ is of order of $\delta^{-2}$, where $\delta = (a/R)^{1/2}\ll 1$ is a formal small parameter of the problem. The wavenumber $k$ and the frequency $\omega\approx kc$ are assigned the order of $\delta^{-1}$. We will clarify our ordering scheme in Sec.~\ref{s3} in more details.

%From the Maxwell equation
From the radial component
    \begin{equation*}
    %\label{maxwell1}
    -\frac{i\omega}{c}\,E_r
    =
    \frac{ikR}{r}\,
    H_z
%    -
%    \parder{H_\theta}{z}
    \end{equation*}
of the Maxwell equation $\rot\vec{H}=\left(1/c\right)\tparder{\vec{E}}{t}$ (here and below we use the Gaussian system of units), we find to the lowest order
    \begin{equation}
    \label{E_r_and_H_r}
    E_r=-H_z\,.
    \end{equation}
This is the relation that holds in a plane electromagnetic wave; in our case it is satisfied approximately to order $\delta^2$.
From the azimuthal component of the same equation
%\note{Note the change: $\frac{1}{r}\parder{rH_z}{r} \to \parder{H_z}{r}$. do we have that error in \cite{Stupakov+2003PhysRevSTAB.6.034401, Stupakov+Phys2009RevSTAB.12.104401}?}
    \begin{equation*}
    %\label{maxwell2}
    -i\,\frac{\omega}{c}\,E_\theta
    =
    -\parder{H_z}{r}
    \,,
    \end{equation*}
we express $E_\theta$ in terms of $H_z$,
    \begin{equation}
    \label{E_theta}
    E_\theta
    =
    -
    %\frac{i}{k}
    \frac{ic}{\omega}
    \parder{H_z}{x}
    \,.
    \end{equation}
%
%\delete{
%    Having expressed $E_\theta$, $H_z$, $H_\theta$, $H_r$ in terms of $E_r$ and $E_z$, we can now derive equations for $E_r$, $E_z$ if we note that $E_z$ and $H_z$ (and hence $E_r$) satisfy the wave equation
%    \begin{gather}\label{wave_equations}
%    \Delta E_z
%    +
%    \frac{\omega^2}{c^2} E_z
%    =0
%    \,,
%    \qquad
%    \Delta E_r
%    +
%    \frac{\omega^2}{c^2} E_r
%    =0
%    \,.
%    \end{gather}
%Let us write the first of Eqs.\ (\ref{wave_equations}) as
%    \begin{gather}\label{wave_equations'}
%    \frac{1}{r}
%    \parder{}{r}
%    r
%    \parder {E_z}{r}
%    +
%    \parder {^2 E_z}{z^2}
%    +
%    \left(
%    \frac{\omega^2}{c^2}
%    -
%    \frac{k^2 R^2}{r^2}
%    \right)
%    E_z
%    =0
%    \,.
%    \end{gather}
%    The same equation holds for $E_r$.
%}
    As to the function $H_{z}(r)$, it obeys the equation
    \begin{gather}\label{wave_equations'}
    \frac{1}{r}
    \parder{}{r}
    r
    \parder {H_z}{r}
    +
    %\parder {^2 H_z}{z^2}
    %+
    \left(
    \frac{\omega^2}{c^2}
    -
    \frac{k^2 R^2}{r^2}
    \right)
    H_z
    =0
    \,.
    \end{gather}%
Substituting $r=R+x$ and expanding in small ratio $x/R$ we find
%\delete{
%    \begin{subequations}
%    \label{master_equations}
%    \begin{align}
%        \Delta_\bot E_z + 2k^2\left(
%            \Lambda+\frac{x}{R}
%        \right) E_z&=0\,
%        ,\\
%        \label{master_equationsB}
%        \Delta_\bot E_r + 2k^2\left(
%            \Lambda+\frac{x}{R}
%        \right) E_r&=0\,
%        ,
%    \end{align}
%    \end{subequations}
%}
    \begin{align}
    \label{master_equations}
        \parder{^{2}H_{z}}{x^{2}}
        +
        %2k^2
        2\frac{\omega^{2}}{c^{2}}
        \left(
            %\Lambda+\frac{x}{R}
            \frac{x}{R}-\Lambda
        \right) H_z&=0\,
        ,
    \end{align}%
where
%\delete{$\Delta_\bot = \partial^2/\partial x^2 + \partial^2/\partial z^2$,}
    \begin{gather}
    \Lambda(\omega,k)
    =
    \frac{1}{2}
    \left(
        %1 - \frac{k^2 c^2}{\omega^2}
        \frac{k^2 c^2}{\omega^2}-1
    \right)
%    \approx
%    \left(
%        %1 - \frac{k c}{\omega}
%        \frac{k c}{\omega}-1
%    \right)
    .
    \end{gather}
Equation~\eqref{master_equations} describes electromagnetic field in the air near the cylinder surface propagating mainly along the axis $\theta$; for this reason it will be referred to as the \emph{paraxial equation} by analogy with the paraxial approximation in geometrical optics.
%
%The parameter $\Lambda$ measures deviation of the phase velocity $\omega/k$ from the speed of light $c$; in what follows we will assume that $\Lambda \sim \delta^2$; this makes all the terms in Eq.~\eqref{master_equations} the same order. Note that in derivation of Eq. (\ref{E_r_and_H_r}) we set $\omega = ck$ which formally corresponds to $\Lambda = 0$; it is easy to check, however, that the equations are still valid for $\Lambda \sim \delta^2$, the error being of order of $\delta^2$.

%\delete{
%    Although we have separate equations (\ref{master_equations}) for $E_r$ and $E_z$, in the next section we will see that the functions $E_r$, $E_z$ are coupled through the boundary  conditions on the conducting wall.
%}

%%%%%%%%%%%%%%%%%%%%%%%%%%%%%%%%%%%
%
%\section{Surface modes propagating in azimuthal direction}
\section{Surface modes}
\label{s3}
%
%%%%%%%%%%%%%%%%%%%%%%%%%%%%%%%%%%%

%For a surface mode propagating in azimuthal direction we set $E_z = H_r =0$ and assume that all other components of the field do not depend on $z$. Note that $H_\theta = 0$ as well. Eq.~\eqref{master_equationsB} can be written as
%    \begin{align}
%        \frac{\p^2 }{\p x^2} H_z + 2k^2\left(
%            \Lambda+\frac{x}{R}
%        \right) H_z&=0
%        ,
%    \end{align}
%and from~\eqref{E_theta_and_H_theta} we find
%    \begin{align}\label{eq:11}
%    E_\theta
%    &=
%    -
%    \frac{i}{k}
%    \frac{\partial H_z}{\partial x}
%    .
%    \end{align}
Eq.~\eqref{master_equations} should be complemented by a boundary condition on the surface of the metal $x=0$. Assuming the skin depth in the metal much smaller than the reduced wavelength $c/\omega$, we use the
Leontovich boundary conditions \cite{Leontovich1944IANUSSR(eng), Leontovich1948(eng), LandauLifshitz1984Pergamon}
    \begin{gather}
    \label{3:1}
    \vec{E}_{t} = \xi \left[\vec{n}\times\vec{B}_t\right]
    \end{gather}
where $\vec n$ is a unit vector normal to the boundary and directed from metal to air, and $\xi$ is a dimensionless surface impedance. The parameter $\xi$ is given by the following formula (see, e.g., \cite[\S59]{LandauLifshitz1984Pergamon})
    \begin{gather*}
    \xi =
    \left(1-i\right)
    \left(\omega/8\pi\sigma\right)^{1/2}
    \equiv
    \xi'-i\xi''
    \end{gather*}
where $\sigma$ is the metal conductivity.
%it is assigned below the order of $\delta$.
Good conductors with $\sigma\gg\omega$ are characterized by a low surface impedance, so we assume that
    \begin{gather}
    \label{4:26a}
    |\xi|\ll 1
    \end{gather}
and assign $\xi$ the order of $\delta$.

The condition \eqref{4:26a} implies for the dielectric permittivity of the metal
    \begin{gather}
    \label{4:26d}
    \varepsilon \approx
    \frac{4\pi i\sigma}{\omega}
    =
    \frac{1}{\xi^{2}}
    \end{gather}
to be large, i.e.
    \begin{gather}
    \label{4:26a'}
    |\varepsilon |\gg 1
    .
    \end{gather}
In this case a general dispersion relation for SW on a planar metal-air interface \cite{Raether1988,Sernelius2001,Maier2007,Klimov2009eng}
    \begin{gather}
    \label{4:33}
        \frac{kc}{\omega}
        =
        \sqrt{\frac{\varepsilon}{1+\varepsilon}}
    \end{gather}
is reduced to
    \begin{gather}
    \label{4:33a}
        \frac{kc}{\omega}
        \approx
        \left(
            1 - \frac{\xi^{2}}{2}
        \right)
        .
    \end{gather}
It assumes for the parameter $\Lambda\approx (kc/\omega-1)$
%in Eq.~\eqref{master_equations}
to be of order of $\delta^{2}$.
It can be seen then from Eq.~\eqref{master_equations} that $x/R$ should also be of order of $\delta^{2}$, as assumed above, for the curvature effect to modify substantially the dispersion relation of the surface waves on a metallic cylinder.

To avoid possible misunderstanding, we note that the condition \eqref{4:26a'} implies that the wave frequency is lower than the plasma frequency, $\omega\ll\omega_{p}$. Thus, a high frequency branch of the surface wave with $\omega \gtrsim \omega _{p}$, which has an electrostatic limit $kc\gg\omega $ at $\varepsilon\to-1$, is outside of the scope of our analysis. As to the low frequency branch, $\omega \ll\omega_{p}$, it is typical for the experiments in the terahertz range of frequencies as at the Novosibirsk Free Electron Laser facility  \cite{Knyazev+2010MST_21_054017, Gerasimov+2012b, Kotelnikov+2013PRA_87_023828}. Practical estimations of the surface impedance for these experiments can be found in Ref. \cite{Gerasimov+2013JOSAB_30_2182}.  Some papers (see eg. \cite{Berry1975JPhysA_8_1952, Polanco+2012PhysLettA_1573, Polanco+OptExpress2014_120}) assumed the Drude expression $\varepsilon =1-\omega_{p}^{2}/\omega^{2}$ for the dielectric permittivity instead of \eqref{4:26d}. Although the Drude approximation is not applicable to realistic metals in the frequency range of interest, we will be able to compare our results against these papers in the limit  $\xi\approx -i/|\varepsilon |^{1/2}$.

It can be deduced from the standard theory of surface electromagnetic waves  \cite{Raether1988,Sernelius2001,Maier2007,Klimov2009eng} that in a planar geometry a low frequency branch of SW evanesces inside the conductor ($x<0$) as $H_{z}\propto\exp(\varkappa_{1}x)$ with
    \begin{gather}
    \label{4:31}
    \varkappa_{1}
    =
    \sqrt{k^{2}-\varepsilon \omega^{2}/c^{2}}
    \approx
    \frac{\omega }{c}\,
    \frac{1}{i\xi}
    = \mathcal{O}(\delta^{-2})
    ,
    \end{gather}
which means that using the Leontovich boundary condition \eqref{3:1} is justified if
    \begin{gather*}
    -\frac{\omega R}{c|\xi|^{2}}\,\im(\xi) \gg 1
    .
    \end{gather*}
Outside of the metal ($x>0$), in the planar geometry $H_{z}\propto\exp(-\varkappa_{2}x)$ with
    \begin{gather}
    \label{4:32}
    \varkappa_{2}
    =
    \sqrt{k^{2}-\omega^{2}/c^{2}}
%    =
%    \frac{\omega }{c}\,
%    \sqrt{\frac{-1}{1+\varepsilon}}
    \approx
    \frac{\omega }{c}\,
    i\xi
    = \mathcal{O}(\delta^{0})
    ,
    \end{gather}
which means that SW is formed if $\xi$ is not purely real.

Looking for a solution of a boundary value problem in the form of surface wave with a given real positive frequency $\omega $, we see that the wavenumber $k$ in Eq.~\eqref{4:33a} should have a positive imaginary part
    \begin{gather}
    \label{4:34}
    \im(k)>0,
    \end{gather}
and the parameter $\varkappa_{2}$ should have a positive real part
    \begin{gather}
    \label{4:34a}
    \re(\varkappa_{2})>0.
    \end{gather}
Any solution that obeys the inequality \eqref{4:34} describes a wave propagating and simultaneously damping in the positive direction of the azimuthal angle $\theta$, and Eq.~\eqref{4:34a} ensures that SW is localized near the conductor surface.
These two conditions mean that physically feasible values of $\xi$ obey the conditions
    \begin{gather}
    \label{4:25a}
    \re(\xi)=\xi'>0,
    \qquad
    \im(\xi)=-\xi''<0
    ,
    \end{gather}
as can be proved independently from a consideration of the energy dissipation inside a conducting body \cite{LandauLifshitz1984Pergamon}.  In what follows we will refer to this property of the surface impedance as
    \begin{gather}
    \label{4:25}
    -\tfrac{1}{2}\pi < \arg(\xi) < 0
    .
    \end{gather}
%We note that by  requesting for $\re(\varkappa_{1})$, $\re(\varkappa_{2})$ and $\im(k)$ to be all positive, one can recover the condition  \eqref{4:25}.

For what follows, it is important to examine at this point the radial dependence of a  planar SW in combination with its temporal dependency:
    \begin{gather*}
    H_{z} \propto \exp
        \left[
            {-\varkappa_{2}x - i\omega t}
        \right]
    .
    \end{gather*}
Putting here explicit expression \eqref{4:32} for $\varkappa_{2}$ yields
    \begin{gather*}
    H_{z} \propto \exp\left[
        -i\omega(\xi'x/c+t)-\omega \xi'' x/c
    \right]
    .
    \end{gather*}
The last term in the square brackets shows that the field is spatially localized near the metal-air interface at $x=0$, whereas the first term describes a wave propagating \emph{towards} the interface from the side of large $x$. We will refer to such SWs as \emph{incoming} surface waves.
%We will argue that previous treatments of SW on a curved metal-air interface were focused on an \emph{outgoing} surface wave, {\color{red}which is far less common \tiny GS: this is not clear}; at least, it is absent in the planar geometry.
Note that all previous treatments of SWs on a curved metal-air interface were focused on the \emph{outgoing} surface waves that are absent in the planar geometry.

%%%%%%%%%%%%%%%%%%%%%%%%%%%%%%%%%%%%
%%
%\section{Solution}
%\label{s4}
%%
%%%%%%%%%%%%%%%%%%%%%%%%%%%%%%%%%%%%

Returning to cylindrical geometry the boundary condition \eqref{3:1} reads
    \begin{align*}
    E_\theta|_{x=0}
    =
    -\xi H_z|_{x=0}
    \,,
    \end{align*}
which using~\eqref{E_theta} reduces to
    \begin{align}\label{eq:13}
    %\frac{i}{k}
    \frac{ic}{\omega}
    \parder{H_z}{x}\bigg|_{x=0}
    =
    \xi H_z\bigg|_{x=0}
    \,.
    \end{align}
To proceed further we denote $k_{0}=\omega /c$ and introduce the dimensionless coordinate
   \begin{align}
    \label{4:1}
    \zeta=\left({2k_{0}^2}/{R}\right)^{1/3}
    x
    \end{align}
and dimensionless parameter
%$\nu $,
    \begin{gather}
    \label{4:2}
    \nu  =
    %-
    \left({2k_{0}^2R^2}\right)^{1/3}\,\Lambda\,
    .
    \end{gather}
This casts the paraxial equation \eqref{master_equations} into the form
    \begin{align}
    \label{4:3}
    %\label{AiryEq}
    \der{^2H_{z}}{\zeta^2} +
    \left(\zeta-\nu \right)H_{z}
    &= 0\,
    ,
    \end{align}
and the boundary condition \eqref{eq:13} at $\zeta=0$ reads
    \begin{align}\label{eq:13-1}
    %\label{4:4}
    \der{H_z}{\zeta}\bigg|_{\zeta=0}
    =
    -i
    %\left(\frac{k_{0}R}{2}\right)^{1/3} \xi
    q
    H_z\bigg|_{\zeta=0}
    \,,
    \end{align}
where
    \begin{gather}
    \label{4:4}
        q = \left({k_{0}R}/{2}\right)^{1/3}\xi
        .
    \end{gather}
We also formulate below an additional boundary condition  when $\zeta \to \infty$. Note that in our ordering scheme all the dimensionless parameters $\zeta$, $\nu $ and $q$ have zero order $\delta^{0}$.

General solution of Eq.~(\ref{4:3}) involves the Airy functions
$\Ai$ and $\Bi$,
    \begin{align}
    \label{4:5}
    %\label{AirySol}
    H_{z}(\zeta) &= C_1\Ai(\nu -\zeta) + C_2\Bi(\nu -\zeta)\,
    ,
    \end{align}
where $C_{1,2}$ are unknown constants. Using asymptotic formulas for large negative values of the argument
%\cite[Eqs.~9.7.9 and~9.7.11]{dlmf.nist.gov/Airy}
\cite[Eqs.~9.7.9 and~9.7.11]{Olver+2010nist},
    \begin{gather}
    \label{4:6}
    \begin{aligned}
    \Ai(-\zeta)
    &\approx
    {\frac{1}{\sqrt{\pi }\,\zeta^{1/4} }
    \sin \left(\frac{2 \zeta
    ^{3/2}}{3}+\frac{\pi }{4}\right)}
    \,,
    \\
    \Bi(-\zeta)
    &\approx
    {\frac{1}{\sqrt{\pi }\,\zeta^{1/4} }
    \cos \left(\frac{2 \zeta
    ^{3/2}}{3}+\frac{\pi }{4}\right)}
    \,,
    \end{aligned}
    \end{gather}
we see that each function contains both outgoing (propagating outward from the cylinder) and incoming (propagating inward from infinity) waves.

We first consider the case with an outgoing wave.
%At first, we will allow an outgoing wave.
Physically it means that a surface wave radiates and its energy is leaking radially to infinity. This same assumption is adopted in Refs.~\cite{Hasegawa+2004ApplPhysLett_84_1835, Hasegawa2007PhysRevA.75.063816}.
%and in the theory of diffraction \cite{Fock1965Pergamon}.
It introduces an extra damping to the wave, which is launched by a source in direction tangent to the surface of the cylinder. It is easy to see that asymptotically
    \begin{align}
    \label{4:7}
    \Ai(\nu-\zeta)
    -i
    \Bi(\nu-\zeta)
    \approx
%    -
%    \frac{(-1)^{3/4}}{\sqrt{\pi }\zeta^{1/4}}
%    \,\e^{\frac{2}{3} i \zeta ^{3/2}}
%    =
    \frac{
        \e^{\frac{2}{3}i (\zeta-\nu)^{3/2}-i\frac{\pi}{4}}
    }{
        \sqrt{\pi}(\zeta-\nu)^{1/4}
    }
%    \approx
%    \frac{
%        \e^{\frac{2}{3}i \zeta^{3/2}-i\frac{\pi}{4}}
%    }{
%        \sqrt{\pi}\zeta^{1/4}
%    }
    \,,
    \end{align}
which (with assumed time dependence $\propto \e^{-i\omega t}$) corresponds to the radial propagation from the cylinder surface to infinity; note that Eqs.~\eqref{4:6} and~\eqref{4:7} are written for the range $-\tfrac{1}{3}\pi < \arg(\zeta-\nu) < \tfrac{1}{3}\pi$. This combination of the Airy functions
    \begin{gather}
    \label{4:8}
    H_{z}(\zeta) = C
    \left[
        \Ai(\nu -\zeta) -i\Bi(\nu -\zeta)
    \right]
    \end{gather}
we will refer to as a solution of the first kind since it can also be expressed through the Airy-Fock function $w_{1}(\nu -\zeta)$ of the first kind (see below).
%; the name will be explained in Sec.~\ref{s7}.
The boundary condition~\eqref{eq:13-1} for this wave reads
    \begin{gather}
    \label{4:11}
    \frac{\Ai'(\nu) -i\Bi'(\nu)}
    {\Ai(\nu) -i\Bi(\nu)}
    -
    i
    %\left(\frac{k_{0}R}{2}\right)^{1/3} \xi
    q
    =0
    \,.
    \end{gather}
It represents a dispersion equation to be resolved regarding $\nu $ at a given parameter $q$. The condition Eq.~\eqref{4:34} restricts a goal range for $\nu $ to upper half of the complex plane, ie.
    \begin{gather}
    \label{4:11a}
        \im(\nu) > 0
        ,
    \end{gather}
and an accurate treatment of the wave \eqref{4:8} at $\zeta\to\infty$ should account for complex nature of $\nu$. This can be accomplished by expanding the asymptote \eqref{4:7} for $|\nu | \ll \zeta$ to the following form
    \begin{gather}
    \label{4:11b}
    \Ai(\nu -\zeta) -i\Bi(\nu -\zeta)
    \approx
    \frac{
        \e^{\frac{2}{3}i \zeta^{3/2}-i\frac{\pi}{4}}
    }{
        \sqrt{\pi}\zeta^{1/4}
    }
    \,
    \e^{-i \nu \zeta^{1/2}}
    ,
    \end{gather}
where the factor $\e^{-i\nu \zeta^{1/2}}$ exponentially grows as long as the condition Eq.~\eqref{4:11a} holds. It means that the solution of the first kind given by Eq.~\eqref{4:8} can hardly describe a surface wave. It would not be surprising since, as we have seen, a planar SW is of the incoming type.

On the contrary, a solution of the second kind,
    \begin{gather}
    \label{4:31a}
    H_{z}(\zeta) = C
    \left[
        \Ai(\nu -\zeta) + i\Bi(\nu -\zeta)
    \right]
    ,
    \end{gather}
describes a radially incoming wave with radially decreasing amplitude at $\zeta\to\infty$, where
    \begin{gather}
    \label{4:11d}
    \Ai(\nu -\zeta) + i\Bi(\nu -\zeta)
    =
    \frac{
        \e^{-\frac{2}{3}i \zeta^{3/2} + i\frac{\pi}{4}}
    }{
        \sqrt{\pi}\zeta^{1/4}
    }
    \e^{i \nu \zeta^{1/2}}
    \end{gather}
and the factor $\e^{i\nu \zeta^{1/2}}$ exponentially vanishes. A corresponding dispersion equation reads
    \begin{gather}
    \label{4:11e}
    \frac{\Ai'(\nu) + i\Bi'(\nu)}
    {\Ai(\nu) + i\Bi(\nu)}
    -
    i
    %\left(\frac{k_{0}R}{2}\right)^{1/3} \xi
    q
    =0
    \,.
    \end{gather}

The combinations of the Airy functions $\Ai$ and $\Bi$ in Eqs. \eqref{4:8} and  \eqref{4:31a} can be recognized as the Airy-Fock functions \cite{Fock1965Pergamon}
    \begin{gather}
    \label{4:41}
    \begin{aligned}
        w_{1}(z)
        &=
        \sqrt{\pi} \e^{i\pi/6}  \Ai\left(\e^{2\pi i/3}z\right)
        ,
        \\
        w_{2}(z)
        &=
        \sqrt{\pi} \e^{-i\pi/6}  \Ai\left(\e^{-2\pi i/3}z\right)
        .
    \end{aligned}
    \end{gather}
It can be checked that
    \begin{gather}
    \label{4:42}
    \begin{aligned}
        w_{1}(z)
        &=
        i\frac{\sqrt{\pi }}{2} \left(\Ai(z)-i \Bi(z)\right)
        ,
        \\
        w_{2}(z)
        &=
        -i\frac{\sqrt{\pi}}{2} \left(\Ai(z)+i \Bi(z)\right)
        .
    \end{aligned}
    \end{gather}
Earlier investigation of Eq.~\eqref{4:11} can be found in \cite{Wait1960AP_8_445, Logan1962AP_10_103}, whereas Eq.~\eqref{4:11e} seems to escape attention of the researchers.

%\section{Wave of the first kind}
\section{Numerical solution}
\label{s4}

\begin{figure*}[!ptbh]
  \centering
  % Requires \usepackage{graphicx}
  %\includegraphics[width=\columnwidth]{ComplexMapAiryDi(R=80,xi=0_3,th=-5pi12)}\\
%  \includegraphics[width=0.33\textwidth]{FigAiryFun2a}\hfill
%  \includegraphics[width=0.33\textwidth]{FigAiryFun2b}\hfill
%  \includegraphics[width=0.33\textwidth]{FigAiryFun2c}\\
%  %\includegraphics[width=0.33\textwidth]{FigAiryFun2d}\\
  \includegraphics[width=\textwidth]{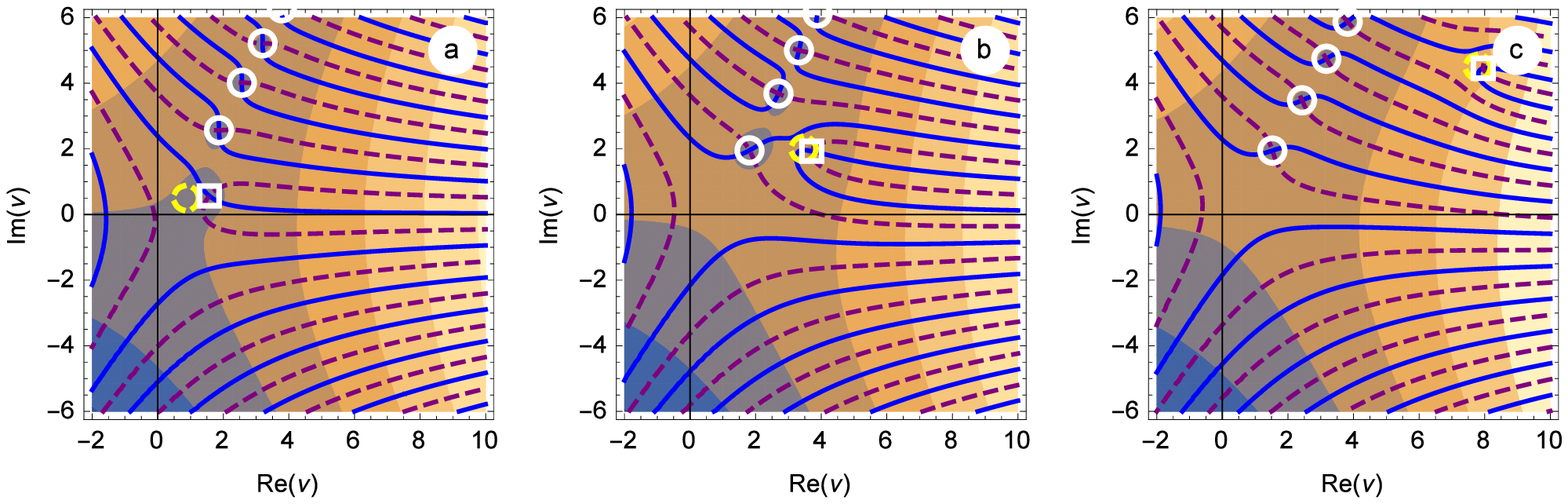}
  \caption{
    (Color online)
    Roots of Eq.~\eqref{4:11} in the complex plane of $\nu $ for various parameter $q=(k_{0}R/2)^{1/3}\xi$:
    (a) $q=1 \e^{-i5\pi/12}$,
        $\nu_\text{PSW}=0.866025 +0.5 i$,  $\nu =1.57272 +0.566718 i$;
    (b) $q=2 \e^{-i5\pi/12}$,
        $\nu_\text{PSW}=3.4641 +2. i$,
        $\nu =3.71063 +1.92959 i$;
    (c) $q=3 \e^{-i5\pi/12}$,
        $\nu_\text{PSW}=7.79423 +4.5 i$,
        $\nu =7.95596 +4.45546 i$;
    %(d) $q=4 \e^{-i5\pi/12}$,
    %    $\nu_\text{PSW}=13.8564 +8. i$,
    %    $\nu =13.9774 +7.96722 i$;
    darker shading corresponds to smaller absolute magnitude of LHS of Eq.~\eqref{4:11}, blue solid and purple dashed curves are the zero levels of real and imaginary part of the LHS, respectively. Yellow dashed circle is centered on the position of the planar SW, white square on the position of cylindrical SW, and white circles are located on the roots that correspond to creeping waves.
  }\label{fig:AiryComplexMap}
%\end{figure*}
%\begin{figure*}[!tbph]
  \medskip
  \centering
%  %\includegraphics[width=\columnwidth]{ComplexMapAiryDi(R=80,xi=0_3,th=-5pi12)}\\
%  \includegraphics[width=0.33\textwidth]{FigAiryFun3a}\hfill
%  \includegraphics[width=0.33\textwidth]{FigAiryFun3b}\hfill
%  \includegraphics[width=0.33\textwidth]{FigAiryFun3c}\\
%  %\includegraphics[width=0.33\textwidth]{FigAiryFun3d}\\
  \includegraphics[width=\textwidth]{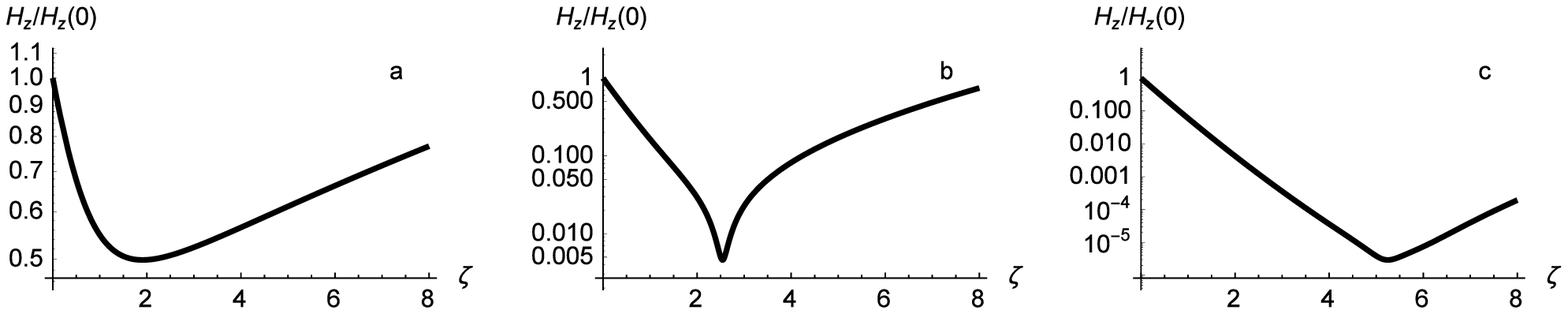}
  \caption{
    %(Color online)
    Radial profiles of the wave amplitudes that correspond to SW roots shown in white squares in Fig.~\ref{fig:AiryComplexMap}:
    (a) $q=1 \e^{-i5\pi/12}$,
        $\nu =1.57272 +0.566718 i$;
    (b) $q=2 \e^{-i5\pi/12}$,
        $\nu =3.71063 +1.92959 i$;
    (c) $q=3 \e^{-i5\pi/12}$,
        $\nu =7.95596 +4.45546 i$;
    %(d) $q=4 \e^{-i5\pi/12}$,
    %    $\nu =13.9774 +7.96722 i$.
    SW wave is formed if $q\gtrapprox 1$, then, the energy flux $S_{r}$ near the metal-air interface reverses inward.
  }\label{fig:AiryRadialProfile}
%\end{figure*}
%\begin{figure*}[!tbph]
  \medskip
  \centering
%  %\includegraphics[width=\columnwidth]{ComplexMapAiryDi(R=80,xi=0_3,th=-5pi12)}\\
%  \includegraphics[width=0.33\textwidth]{FigAiryFun4a}\hfill
%  \includegraphics[width=0.33\textwidth]{FigAiryFun4b}\hfill
%  \includegraphics[width=0.33\textwidth]{FigAiryFun4c}\\
%  %\includegraphics[width=0.33\textwidth]{FigAiryFun3d}\\
  \includegraphics[width=\textwidth]{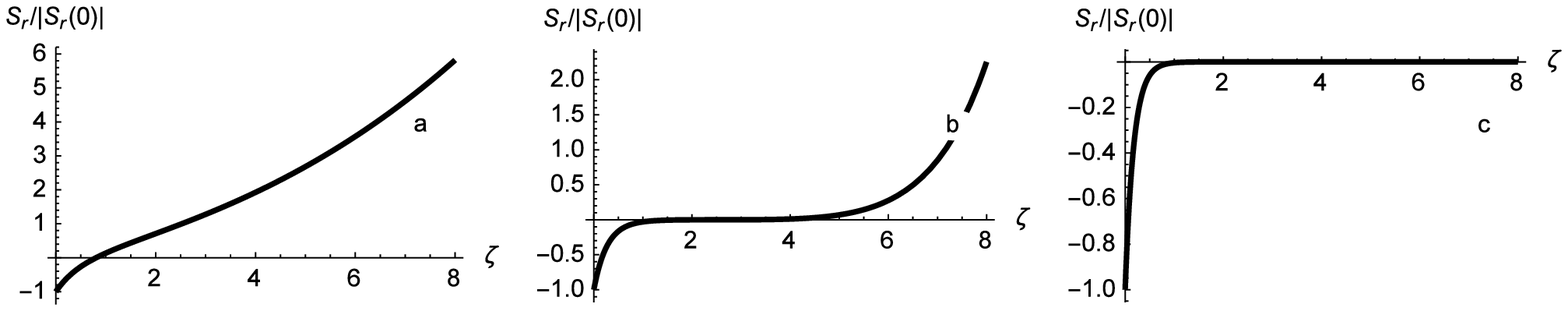}
  \caption{
    %(Color online)
    Radial energy flux that correspond to SW roots shown in white squares in Fig.~\ref{fig:AiryComplexMap}:
    (a) $q=1 \e^{-i5\pi/12}$,
        $\nu =1.57272 +0.566718 i$;
    (b) $q=2 \e^{-i5\pi/12}$,
        $\nu =3.71063 +1.92959 i$;
    (c) $q=3 \e^{-i5\pi/12}$,
        $\nu =7.95596 +4.45546 i$;
    %(d) $q=4 \e^{-i5\pi/12}$,
    %    $\nu =13.9774 +7.96722 i$.
    SW wave is formed if $q\gtrapprox 1$, then, the energy flux $S_{r}$ near the metal-air interface reverses inward from outward as in Figs.~(a) and (b).
  }\label{fig:AiryPoynting}
\end{figure*}

Fig.~\ref{fig:AiryComplexMap} illustrates the results of numerical search of complex roots of the dispersion equation~\eqref{4:11} for the wave of the first kind. Using Wolfram Mathematica \cite{www.wolfram.com} to calculate the Airy  functions, the roots of Eq.~\eqref{4:11} can be found numerically. They are located on the intersections of the zeroth isolevels of the real and imaginary parts of LHS of Eq.~\eqref{4:11} (multiplied by the denominator of the first term).  These isolevels are represented in Figs.~\ref{fig:AiryComplexMap} by the blue (solid) and purple (dashed) lines, respectively.
%\note{
%        Because of the specific form of Eq.~\protect\eqref{4:11} zeros of the denominator in the first term are also located on the intersection of these line and must therefore be disregarded from actual roots of Eq.~~\eqref{4:11}.
%}

%Inspecting these figures shows that Eq.\eqref{4:11} has many roots. In particular, a (presumably) infinite series of roots is located near the ray $\arg(\nu)=\tfrac{1}{3}\pi$ (they are marked by white circles). However, the radial profiles of $H_{z}(\zeta)$, which correspond to these roots, have nothing in common with those expected for SW since $H_{z}(\zeta)$ grows with $\zeta$. Similar roots, numerically found in Ref.~\cite{Liaw+2008OE.16.4945} from Hasegawa's dispersion equation \cite{Hasegawa+2004ApplPhysLett_84_1835}, were erroneously reported as belonging to SWs. We identify these roots as creeping modes. True roots, which correspond to SW, are marked by white squares;  they are separated from the creeping modes in Figs.~\ref{fig:AiryComplexMap}c and \ref{fig:AiryComplexMap}d.

Inspecting these figures shows that Eq.~\eqref{4:11} has many roots. In particular, a (presumably) infinite series of roots is located near the ray $\arg(\nu)=\tfrac{1}{3}\pi$.
We found that the roots marked by the white circles in Fig.~\ref{fig:AiryComplexMap} have radial profiles of $H_{z}(\zeta)$ monotonically growing with $\zeta$ as expected for the creeping waves (see discussion of CW in the Introduction). Hence we associate these roots with the creeping waves.
% As we mentioned in the Introduction, the term ``creeping wave''  is used in modern theory of diffraction \cite{Honl1961Springer, Rulf1967JMP_8_1785, Rulf1967JMP_8_1785, PaknysWang1986AP_34_674, Andronov+1995JEWA_9_503, PaknysJackson2005IEEETransAP_53_898, JosefssonPersson2006Conformal, Andronov2008AJ_54_139} for a refracted wave on convex surfaces, where an extra attenuation due to the surface curvature is present. As explained in Ref.~\cite[chapt.~5]{Honl1961Springer}, the creeping waves rapidly damp in the shadow region created by a convex obstacle on the way of a wave propagation. They damp both in the direction of propagation and towards to the surface of the obstacle. In other words, the amplitude of a creeping wave grows with the distance along the normal to the surface from the shadow region to the lit one.

%Similar roots, numerically found in Ref.~\cite{Liaw+2008OE.16.4945} from Hasegawa's dispersion equation \cite{Hasegawa+2004ApplPhysLett_84_1835}, were erroneously reported as belonging to SWs. We identify these roots as creeping modes. True roots, which correspond to SW, are marked by white squares;  they are separated from the creeping modes in Figs.~\ref{fig:AiryComplexMap}c and \ref{fig:AiryComplexMap}d.

The only exception from monotonically growing modes is represented by the root located at the center of the white square in Fig.~\ref{fig:AiryComplexMap}a, b, and c. The radial profiles of the modes, corresponding to this root are plotted in Fig.~\ref{fig:AiryRadialProfile}a, b, and c, respectively for 3 different values of the parameter $q$. Comparing these plots, we see that with increasing $q$ the radial profile of these modes is gradually converted to a radially decaying pattern, which is expected for a surface mode localized in the vicinity of a metal-air interface. Such a localized mode we identify with a surface electromagnetic wave (SW).

It is important to note, that no SWs were found if $\arg(\xi)>-\tfrac{1}{3}\pi$. In the case, where $-\tfrac{1}{2}\pi <\arg(\xi) < -\tfrac{1}{3}\pi$, a single surface-like mode is formed near the metal-air interface for every given value of $q$ if $|q|\gtrsim 1$. For a small $q$, an amplitude of such a mode initially diminishes but eventually becomes to grow at larger $\zeta$ as shown in Fig.~\ref{fig:AiryRadialProfile},a. We note also, that the SW root is visibly well separated from the other roots in Fig.~\ref{fig:AiryComplexMap} when SW becomes radially localized (as in Fig.~\ref{fig:AiryComplexMap},b,c), whereas it is located in the vicinity of the other roots in the case if SW is not well localized (as in Fig.~\ref{fig:AiryComplexMap},a). A similar behavior of the SW roots was noted in Ref.~\cite{PaknysJackson2005IEEETransAP_53_898}. It was stated there that the SW roots freely migrate in the complex map of the parameter $n=kR$, whereas the CW roots remain tightly confined to the immediate
vicinity of the Stokes line of the Hankel functions.

The radial energy flux given by the radial component $S_{r} \propto \re(E_{\theta}^{\ast}H_{z})$ of the Poynting vector for the SW is shown in Fig.~\ref{fig:AiryPoynting}. It is directed inward to the metal-air interface in its vicinity but is reversed outward at larger distances $\zeta \gtrsim |\nu |$, where it might be too small as in Fig.~\ref{fig:AiryPoynting},c. The change of sign of $S_{r}$ is a manifestation of the two-dimensional nature of the wave which propagates mainly along the metal surface but also damps in that direction because of the energy leakage both into the metal and radially to the infinity.

\begin{figure*}[!ptbh]
  \centering
%  \includegraphics[width=0.33\textwidth]{FigAiryFun6a}\hfill
%  \includegraphics[width=0.33\textwidth]{FigAiryFun6b}\hfill
%  \includegraphics[width=0.33\textwidth]{FigAiryFun6c}\\
%  %\includegraphics[width=0.33\columnwidth]{FigAiryFun6d}\\
  \includegraphics[width=\textwidth]{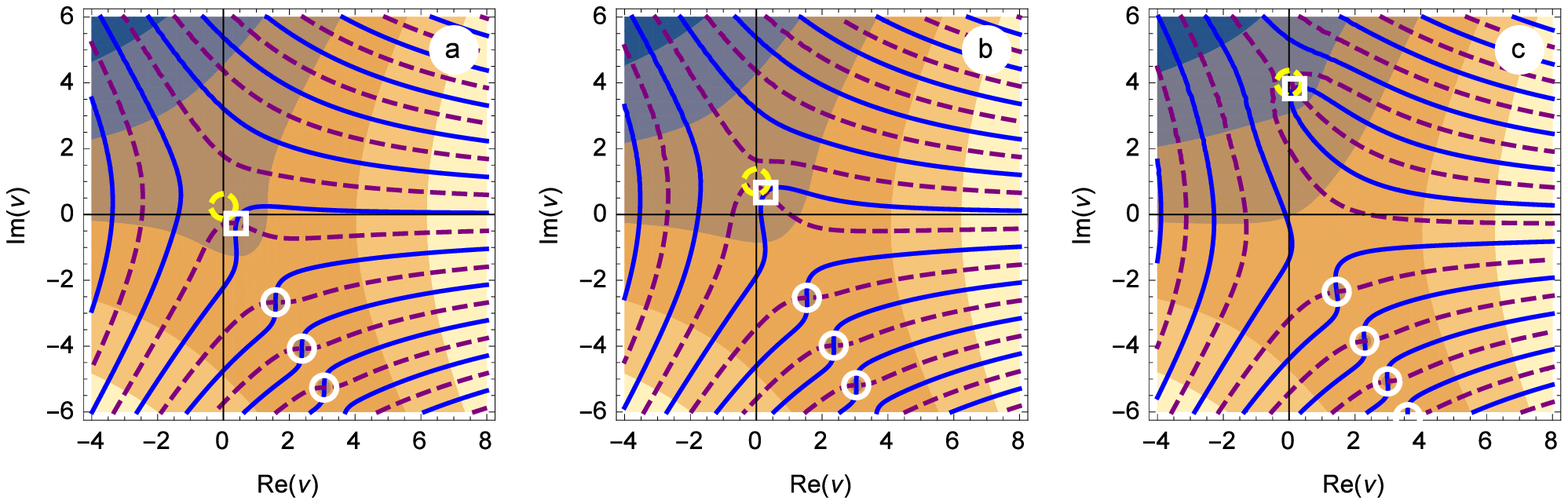}
  \caption{
    (Color online)
    Roots of Eq.~\eqref{4:11e} in the complex plane of $\nu $ for various values of $q$:
    (a) $q=0.5 \e^{-i\pi/4}$,
        $\nu_\text{PSW}=0.25i$,
        $\nu = 0.386713 -0.273652 i$;
    (b) $q=\e^{-i\pi/4}$,
        $\nu_\text{PSW}=1i$,
        $\nu = 0.286572 +0.667116 i$;
    (c) $q=2 \e^{-i\pi/4}$,
        $\nu_\text{PSW}=4i$,
        $\nu = 0.169778 +3.82383 i$;
    %(d) $q=3 \e^{-i\pi/4}$,
    %    $\nu_\text{PSW}=9i$,
    %    $\nu = 0.116359 +8.8822 i$;
    darker shading corresponds to smaller absolute magnitude of LHS of Eq.~\eqref{4:11e}, blue solid and purple dashed curves are the zero levels of real and imaginary part of the LHS, respectively. Yellow dashed circle is centered on the position of the planar SW, white squares on cylindrical SW, and white circles are located on the roots that correspond to creeping waves.
  }\label{fig:AiryComplexMapPlus}
%\end{figure*}
%\begin{figure*}[!ptbh]
  \medskip
  \centering
%  \includegraphics[width=0.33\textwidth]{FigAiryFun7a}\hfill
%  \includegraphics[width=0.33\textwidth]{FigAiryFun7b}\hfill
%  \includegraphics[width=0.33\textwidth]{FigAiryFun7c}\\
%  %\includegraphics[width=\columnwidth]{FigAiryFun7d}\\
  \includegraphics[width=\textwidth]{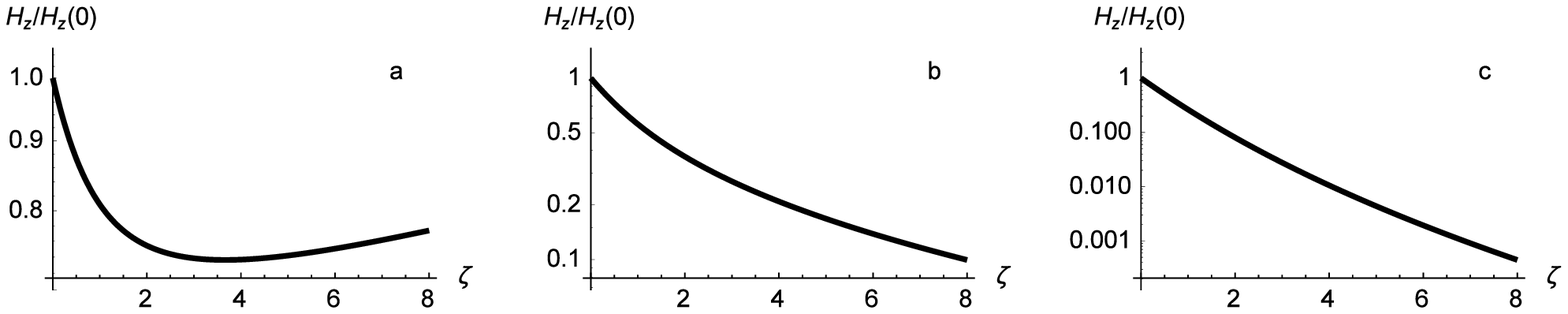}
  \caption{
    %(Color online)
    Radial profiles of the amplitudes that correspond to SW roots shown in white squares in Fig.~\ref{fig:AiryComplexMapPlus}:
    (a) $q=0.5 \e^{-i\pi/4}$,
        %$\nu_\text{PSW}=0.25i$,
        $\nu = 0.386713 -0.273652 i$;
    (b) $q=\e^{-i\pi/4}$,
        %$\nu_\text{PSW}=1i$,
        $\nu = 0.286572 +0.667116 i$;
    (c) $q=2 \e^{-i\pi/4}$,
        %$\nu_\text{PSW}=4i$,
        $\nu = 0.169778 +3.82383 i$;
    %(d) $q=3 \e^{-i\pi/4}$,
    %    %$\nu_\text{PSW}=9i$,
    %    $\nu = 0.116359 +8.8822 i$;
    SW wave is formed if $q\gtrapprox 1$, then $\im(\nu)>0$ and the wave amplitude exponentially decreases with $\zeta$.
  }\label{fig:AiryRadialProfilePlus}
%\end{figure*}
%\begin{figure*}[!ptbh]
  \medskip
  \centering
%  \includegraphics[width=0.33\textwidth]{FigAiryFun8a}\hfill
%  \includegraphics[width=0.33\textwidth]{FigAiryFun8b}\hfill
%  \includegraphics[width=0.33\textwidth]{FigAiryFun8c}\\
%  %\includegraphics[width=\columnwidth]{FigAiryFun78d}\\
  \includegraphics[width=\textwidth]{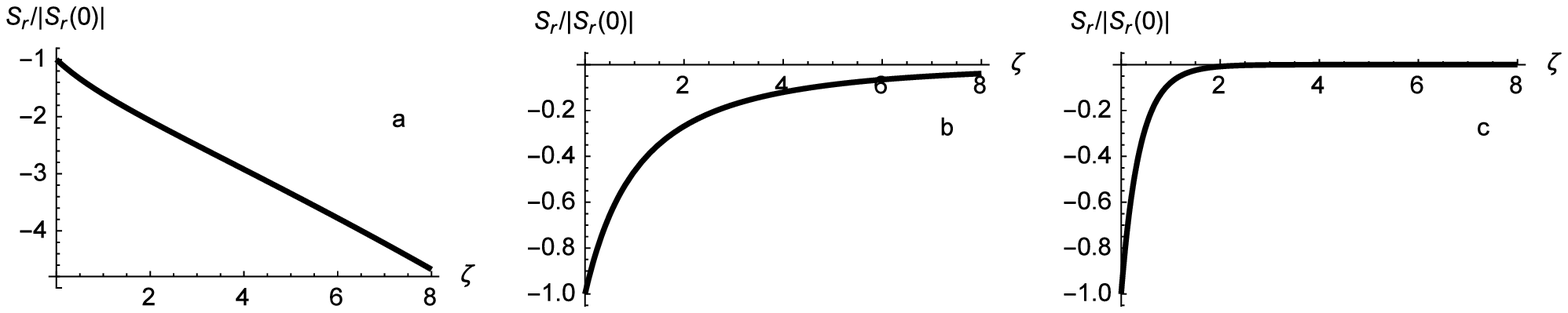}
  \caption{
    Radial energy flux that correspond to SW roots shown in white squares in Fig.~\ref{fig:AiryComplexMapPlus}:
    (a) $q=0.5 \e^{-i\pi/4}$,
        %$\nu_\text{PSW}=0.25i$,
        $\nu = 0.386713 -0.273652 i$;
    (b) $q=\e^{-i\pi/4}$,
        %$\nu_\text{PSW}=1i$,
        $\nu = 0.286572 +0.667116 i$;
    (c) $q=2 \e^{-i\pi/4}$,
        %$\nu_\text{PSW}=4i$,
        $\nu = 0.169778 +3.82383 i$;
    %(d) $q=3 \e^{-i\pi/4}$,
    %    %$\nu_\text{PSW}=9i$,
    %    $\nu = 0.116359 +8.8822 i$;
    SW wave is formed if $q\gtrapprox 1$; the energy flux $S_{r}$ everywhere is directed inward.
  }\label{fig:AiryPoyntingPlus}
\end{figure*}
As to the wave of the second kind, Eq.~\eqref{4:11e} has a SW-like solution for the entire allowed range $-\tfrac{1}{2}\pi < \arg(\xi) < 0$ provided that $|q| \gtrsim 1$. The results of numerical computations are shown in Figs.~\ref{fig:AiryComplexMapPlus}, \ref{fig:AiryRadialProfilePlus}, \ref{fig:AiryPoyntingPlus}. Serial roots, which might also be thought of as creeping waves, are located near the ray $\arg(\nu)=-\tfrac{1}{3}\pi$ as shown Fig.~\ref{fig:AiryComplexMapPlus}.
Note however that $\im(k)<0$ for these roots and, hence, corresponding modes amplify in the direction of propagation along the metal surface due to energy influx from large radii.
Alone root, which corresponds to SW, is well separated from the CW roots. It passes to the upper half of the complex plane $\im(\nu)>0$ if $|q|\gtrsim 1$ but is located in the lower half-plane $\im(\nu)<0$ if $|q|\ll 1$ as shown in Fig.~\ref{fig:AiryComplexMapPlus},a. In the latter case, the wave amplitude eventually grows with $\zeta$ as in Fig.~\ref{fig:AiryRadialProfilePlus},a.
%In case of $|q|\gtrsim$ the radial profile of $|H_{z}(\zeta)|$  monotonically decreases to zero as shown in Figs.~\ref{fig:AiryRadialProfilePlus}b,c, and the energy radial flux $S_{r}$ is everywhere directed inwards as in  \ref{fig:AiryPoyntingPlus}b,c.
%
In case of $|q|\gtrsim 1$ the radial profile of $|H_{z}(\zeta)|$  monotonically decreases to zero as shown in Figs.~\ref{fig:AiryRadialProfilePlus},b,c. The energy radial flux $S_{r}$ is everywhere directed inwards as shown in Fig.~\ref{fig:AiryPoyntingPlus}. This fact distinquishes the surface wave of second type from the first one. Both types are characterized by inward radial energy flux in the vicinity of the metal-air interface as could be expected for a realistic metallic cylinder that should dissipate the wave energy, however at large distances the radial energy flux in the wave of the first type reverses its direction. For this reason we call the surface wave of the first type also the outgoing wave whereas the surface wave of the second type is referred to as the incoming wave.

%The radial energy flux given by the radial component $S_{r} = \left(c/4\pi\right) E_{\theta}H_{z}$ of the Poynting vector is shown in Fig.~\ref{fig:AiryPoynting}. It is directed inward to the metal-air interface in its vicinity but is also reversed outward at larger distances $\zeta \gtrsim |\nu |$, however it might be too small there as in Fig.~\ref{fig:AiryPoynting}c.

We note also that the radial profile of CWs (not shown in Fig.~\ref{fig:AiryRadialProfilePlus}) exhibits growth in the vicinity of the metal-air interface in contrast to the surface waves.
%We found at most by one surface mode in both types for any value of $q$.
We found that for any value of $q$ there is at most one surface mode of both types.

% Thus, we see that inspecting the behavior of the Hankel functions at $r\to\infty$ in earlier theories of the surface waves

%\section{Relation to the surface waves in planar geometry}
\section{Analytical solution}
\label{s5}

Let us see how we can recover a SW on a planar surface in the limit $R\to \infty$. For a large $R$ the parameter $q$ also becomes large, $q\gg 1$, and as we will see from the result, $|\nu| \gg 1$. We start from Eq.~\eqref{4:11} for the wave of first kind and use the asymptotic expansion
%\note{The asymptotic formula \eqref{4:12} is valid in . {\color{red} Is this range correct? I would expect $-\tfrac{1}{3}\pi< \arg(\nu)<\tfrac{1}{3}\pi$.}
%} See
%\cite[Eqs.~9.7.5 and~9.7.13]{dlmf.nist.gov/Airy}
\cite[Eqs.~9.7.5 and~9.7.13]{Olver+2010nist}

    \begin{align}
    \label{4:12}
    \Ai(\nu)
    -i
    \Bi(\nu)
    \approx
    -
    \frac{
        i\,\e^{\frac{2}{3} \nu ^{3/2}}
    }{
        \sqrt{\pi}\nu ^{1/4}
    }
    \end{align}
valid in the range $-\pi< \arg(\nu)<\tfrac{1}{3}\pi$ and $|\nu| \gg 1$. This  reduces Eq.~\eqref{4:11} to
    \begin{align}
    %\label{eq:13-3}
    \label{4:14}
    \sqrt{\nu }
    =
    i
    %\left({k_{0}R}/{2}\right)^{1/3} \xi
    q
    \,.
    \end{align}
Squaring this equation and using~\eqref{4:2} and~\eqref{4:4} we obtain
    \begin{align}%\label{eq:13-4}
    \label{4:16}
    \Lambda
    =
    -
    \frac{1}{2}
    \xi^2
    \,,
    \end{align}
which can readily be recognized as the dispersion relation \eqref{4:33a} of a planar surface wave (PSW).
%Indeed, assuming that the dielectric permittivity of the metal $\varepsilon\approx 1/\xi^{2} $ is large by absolute magnitude, $|\varepsilon |\gg 1$, the standard dispersion expression for the planar surface wave can be cast into the form
%    \begin{gather}
%    \label{4:20}
%    \frac{kc}{\omega}
%    =
%    \sqrt{\frac{1}{1+\varepsilon}}
%    \approx
%    1-\frac{\xi^{2}}{2}
%    ,
%    \end{gather}
%which is actually the same as Eq.~\eqref{4:16} since $\Lambda\approx 1-kc/\omega$.

Having obtained the solution~\eqref{4:14} we need to verify if the argument of $\nu $ lies within the intersection of the range that was assumed in the asymptotic expansion~\eqref{4:12} and the target range \eqref{4:11a}. It follows from~\eqref{4:14} that $\arg \nu  = 2(\arg\xi + \pi/2)$ and one finds that this is indeed the case if $-\tfrac{1}{2}\pi < \arg\xi < -\tfrac{1}{3}\pi$ (here we take into account that the argument of $\xi$ lies in the interval given by~\eqref{4:25}).

Radial profile of PSW near the metal-air interface can be recovered using the asymptotic expression \eqref{4:12} and expanding the function \eqref{4:8} for $\zeta\ll|\nu |$:
    \begin{gather}
    \label{4:22a}
    H_{z}(\zeta)
    \approx
    -
    C
    \frac{
        i\,\e^{\frac{2}{3} \nu ^{3/2}}
    }{
        \sqrt{\pi}\nu ^{1/4}
    }\,
    \e^{ \sqrt{\nu }\,\zeta }
    \propto
    \e^{ i \xi\,k_{0}\,x }
    .
    \end{gather}
The condition $\zeta\ll|\nu |$ means that the surface wave is planar at the distance
    \begin{gather}
    \label{4:23}
        x\ll x_{0}
        =
        \left|\xi\right|^{2}R/2
    \end{gather}
from the metal-air interface. At larger distances, $\zeta\gg|\nu |$, it is transformed into an outgoing wave \eqref{4:11b} with exponentially growing amplitude as is seen in Figs~\ref{fig:AiryRadialProfile}a-c.

To obtain a radiative correction to the dispersion relation~\eqref{4:16} of PSW, one needs to take a more accurate asymptotic expression for the Airy functions. Although $\Ai(\nu)$ and $\Bi(\nu)$ are entire functions analytical in the entire complex plane of their argument without any cuts, they have different asymptotic expressions in various segments of the complex plane which are separated by Stokes' lines $\arg(\nu)=\pm \tfrac{1}{3}\pi$ and
%$\arg(\zeta)=\pm 3\pi/3$,
$\arg(\nu)=\pm\pi$.
%, and $\arg(\zeta)=0$.
In particular, as it has been already mentioned, the asymptotic expression Eq.~\eqref{4:12} is valid if $-\pi<\arg(\nu)<\tfrac{1}{3}\pi$. Recall that we are looking for a solution with $\nu$ in the range $0\leq\arg(\nu)\leq \pi$ which describes the waves propagating and damping in the positive direction of the angle coordinate $R\theta$. An appropriate asymptote for this case
%\cite[Eqs.~9.7.5 and~9.7.13]{dlmf.nist.gov/Airy}
\cite[Eqs.~9.7.5 and~9.7.13]{Olver+2010nist}
    \begin{gather}
    \label{4:21}
    \Ai(\nu)-i\Bi(\nu)
    \approx
    -
    \frac{
        i\,\e^{\frac{2}{3} \nu ^{3/2}}
    }{
        \sqrt{\pi}\,\nu ^{1/4}
    }
    +
    \frac{
        \e^{-\frac{2}{3} \nu ^{3/2}}
    }{
        \sqrt{\pi}\,\nu ^{1/4}
    }
    \end{gather}
contains an additional term as compared to Eq.~\eqref{4:12} and is appropriate for the range $-\tfrac{1}{3}\pi < \arg(\nu) < \pi$. The second term in \eqref{4:21} is negligible (subdominant) if $-\tfrac{1}{3}\pi < \arg(\nu) < \tfrac{1}{3}\pi$ but it dominates if $\arg(\nu)$ exceeds $\tfrac{1}{3}\pi$.  Putting \eqref{4:21} into Eq.~\eqref{4:11} casts it to the form
    \begin{gather}
    \label{4:22}
    \e^{\frac{4}{3} \nu ^{3/2}}
    =
    -
    \frac{
        %\sqrt{\nu } + i(k_{0}R/2)^{1/3}\xi
        \sqrt{\nu } + iq
    }{
        %i\sqrt{\nu } + (k_{0}R/2)^{1/3}\xi
        i\sqrt{\nu } + q
    }
    \,.
    \end{gather}
If $|\arg(\nu)|<\tfrac{1}{3}\pi$ and $|\nu |\gg 1$, the left-hand side (LHS) of this equation is large, which means that the denominator of the right-hand side (RHS) tends to zero, thus reproducing Eq.~\eqref{4:14}. We now take $\sqrt{\nu }$ from Eq.~\eqref{4:14} as a first approximation and put it back into Eq.~\eqref{4:22} everywhere except for the denominator in RHS. This yields a corrected expression
    \begin{gather}
    \label{4:24}
    \Lambda
    =
    -
    %\left(
    %    k_{0}R/2
    %\right)^{1/3}
    \frac{\xi^{2}}{2}
    \left[
        1
        -
        4i\e^{-\frac{2}{3}k_{0}R\left(i\xi\right)^{3}}
    \right]
    ,
    \end{gather}
where the second term stands for the radiative addition to the outgoing SW dispersion relation. It is responsible for additional damping of SW, which is characterized by $\im(k)=k_{0}\im(\Lambda)$. In particular, a planar surface wave does not damp in the direction of propagation in case of pure imaginary surface impedance, $\arg(\xi)=-\tfrac{1}{2}\pi$, since the first term in \eqref{4:24} turns out to be purely real. However, the second term is imaginary giving rise to radiative damping of the surface wave. According to \eqref{4:24}, this sort of damping becomes exponentially weak as the curvature radius $R$ tends to infinity. This result was earlier obtained for a particular case of purely imaginary (inductive) surface impedance
by M.~Berry \cite{Berry1975JPhysA_8_1952} (with tow times smaller coefficient before exponential term in his Eq.~28) and by N.~Logan and K.~Yee in Ref.~\cite{Logan1962AP_10_103} (see their unnumbered equation after Eq.~12).

The effect of the curvature of the metal-air interface on the wave of the first kind becomes much stronger if $\xi$ acquires a real part. Depending on the value of $\arg(\xi)$ the second term in Eq.~\eqref{4:24} can be either decreasing or growing function of $R$. When $\arg(\xi)$ approaches or exceeds the value $-\tfrac{1}{3}\pi$, the second term in Eq.~\eqref{4:24} grows exponentially with $R$ and for large values of $R$ exceeds the first one. Since we used a perturbation theory, which assumes that the second term should be small in comparison with the first one, Eq.~\eqref{4:24} becomes invalid in this case. Our numerical calculations indicate that Eq.~\eqref{4:11} has no solution in the form of SW if
    \begin{gather}
    \label{4:26b}
    \arg(\xi) > - \frac{\pi}{3}
    .
    \end{gather}
Thus, an outgoing SW does not exist for all physically reliable values of the surface impedance $\xi$ even in the limit of large radius $R$ of conducting cylinder.

Numerical study of the problem detailed in Sec.~\ref{s4} showed that Eq.~\eqref{4:11} has a large (perhaps, infinite) number of solutions which exhibit an amplitude growing with the distance from metal-air boundary. Few such lowest  modes were found numerically in Ref.~\cite{Liaw+2008OE.16.4945} but they were erroneously identified as surface modes. We argue that these are the creeping modes. To find their dispersion relation, we assume that either %$|\sqrt{\nu }|\gg (2k_{0}^{2}R^{2})^{1/3}|\xi|$ or %$|\sqrt{\nu }|\ll (2k_{0}^{2}R^{2})^{1/3}|\xi|$.
$|\sqrt{\nu }|\gg |q|$ or
$|\sqrt{\nu }|\ll |q|$.
Then Eq.~\eqref{4:22} reduces to
    \begin{gather}
    \label{4:26c}
    \e^{\frac{4}{3} \nu ^{3/2}}
    =
    \pm
    i
    \,,
    \end{gather}
%%
%% Эти формулы есть в книге \cite{Honl+1964Mir}, стр. 352, Eq.141.1.
%%
where the upper sign corresponds to the case $|\sqrt{\nu }|\gg |q|$, and the lower sign to $|\sqrt{\nu }|\ll |q|$. Noting that $\pm i=\e^{\pm i\pi/2+i2\pi i j}$, where integer number $j$ runs from $0$ to $\infty$, we obtain an infinite number of roots,
    \begin{gather}
    \label{4:27}
%        \Lambda_{j}
%        =
%        -
%        \left(3/4\right)^{2/3}
%        \left(\pm\pi/2+2\pi j\right)^{2/3}
%        \e^{i\pi/3}
%        \left({2k_{0}^2R^2}\right)^{-1/3}
%        .
        \Lambda_{j}
        =
        %-
        \left(3\pi/2\right)^{2/3}
        \left(j\pm1/4\right)^{2/3}
        \e^{i\pi/3}
        \left({2k_{0}^2R^2}\right)^{-1/3}
        .
    \end{gather}
In Figs.~\ref{fig:AiryComplexMap}, these roots are located on the ray $\arg(\nu)=\tfrac{1}{3}\pi$ and a few lowest roots are encircled by white circumferences. To the best of our knowledge, the dispersion relation \eqref{4:27} was first derived in Ref.~\cite[Eq.~141.1]{Honl1961Springer}.

%%%%%%%%%%%%%%%%%%%%%%%%%%%%%%%%%%%
%
%\section{Incident surface wave}
%\label{s6}

Let us now consider the wave of the second kind which is described by Eq.~\eqref{4:31a} and obeys the dispersion equation Eq.~\eqref{4:11e}. As discussed in Sec.~\ref{s3}, this wave, if exists, could be evanescent at $\zeta\gg|\nu |$ according to  Eq.~\eqref{4:11d} provided that Eq.~\eqref{4:11a} holds.

In the range $-\tfrac{1}{3}\pi < \arg(\nu) < \pi$ an appropriate asymptotes at for $|\nu |\to\infty$ have the form
%\cite[Eqs.~9.7.5 and~9.7.5, 9.7.6, 9.7.13, and 9.7.14]{dlmf.nist.gov/Airy}
\cite[Eqs.~9.7.5 and~9.7.5, 9.7.6, 9.7.13, and 9.7.14]{Olver+2010nist}
    \begin{gather}
    \label{6:12}
    \begin{aligned}
    \Ai(\nu)
    +i
    \Bi(\nu)
    &\approx
    \frac{
        i\,\e^{\frac{2}{3} \nu ^{3/2}}
    }{
        \sqrt{\pi}
    }
    \nu ^{-1/4}
    \left(
        1
        +
        \frac{5}{48 \nu ^{3/2}}
    \right)
    ,\\
    \Ai'(\nu)
    +i
    \Bi'(\nu)
    &\approx
    \frac{
        i\,\e^{\frac{2}{3} \nu ^{3/2}}
    }{
        \sqrt{\pi}
    }
    \nu ^{1/4}
     \left(
        1
        -
        \frac{7}{48 \nu ^{3/2}}
    \right)
    ;
    \end{aligned}
    \end{gather}
note the absence of subdominant terms proportional to $\exp\left({-\tfrac{2}{3} \nu ^{3/2}}\right)$. Putting these asymptotes into Eq.~\eqref{4:11e} yields the equation
    \begin{align}
    \label{6:14}
    \sqrt{\nu }
    \left(
        1 - \frac{1}{4\nu ^{3/2}}
    \right)
    =
    i
    %\left({k_{0}R}/{2}\right)^{1/3} \xi
    q
    .
    \end{align}
Solving it by iterations for the case $|\nu |\gg1$, we find
    \begin{align}
    \label{6:16}
    \Lambda
    =
    -
    \frac{\xi^{2}}{2}
    %+
    -
    \frac{i}{2k_{0}R\xi}\,
    ,
    \end{align}
where the second term in RHS again represents a radiative correction to the dispersion of planar SW.
%In original notations, Eq.~\eqref{6:16} reads
%    \begin{gather}
%    \label{6:17}
%    \frac{kc}{\omega}
%    =
%    1
%    - \frac{\xi^{2}}{2}
%    -
%    \frac{i}{2k_{0}R\xi}
%    .
%    \end{gather}
%It is valid if $|q|=\left(k_{0}R/2\right)^{1/3}|\xi| \gg 1$, and the last term is small in comparison to the other two.
Eq.~\eqref{6:16} is valid if $|q|=\left(k_{0}R/2\right)^{1/3}|\xi| \gg 1$, and the last term is small in comparison to the first one. It means that a surface wave of the second kind exists for any $\xi$ in the allowed range \eqref{4:25} if $|q| \gtrsim 1$. This conclusion is supported by numerical simulations as reported in Sec.~\ref{s4}. It is also found that SW of the second kind disappears for $|q|\lesssim1$ since $\im\nu <0$  and SW mode is transformed into a radially growing mode. A critical value of $q$ can be in principle found for which $\im\nu = 0$, and it occurred to be a function of $\arg(\xi)$.

It is somewhat surprising that the last term in \eqref{6:16}, which stands for the radiative correction to the dispersion relation of SW, diminishes $\im(k)=k_{0}\Lambda$. It may reverse the sign of $\im(k)$ if $-\im(\xi) <1/(2k_{0}R|\xi|^{2})$ and this conclusion is supported by numerical simulations: compare $\nu$ and $\nu_\text{PSW}$ in the  caption to Fig.~\ref{fig:AiryComplexMapPlus}.

It is worth noting that a result similar to Eq.~\eqref{6:16} was obtained by J. Wait \cite{Wait1960AP_8_445} although it is not clear which of the two equations \eqref{4:11}  and \eqref{4:11e} he used.

In conclusion to this section, we note that serial roots are located near the ray $\arg(\nu)=-\tfrac{1}{3}\pi$ and they are complex conjugate to similar roots \eqref{4:27} found for the wave of the first kind.

\section{Concave case}
\label{s6}
%
%%%%%%%%%%%%%%%%%%%%%%%%%%%%%%%%%%%

Reversing the sign of $x$ in Eq.~\eqref{master_equations} yields
    \begin{align}
    \label{5:1}
        \parder{^{2}H_{z}}{x^{2}}
        -
        2\frac{\omega^{2}}{c^{2}}
        \left(
            \frac{x}{R}
            +
            \Lambda
        \right) H_z&=0\,
        .
    \end{align}%
Keeping the same notations \eqref{4:1} and \eqref{4:2} as for the convex case brings the last equation to the form
    \begin{align}
    \label{5:3}
    \der{^2H_{z}}{\zeta^2}
    -
    \left(\zeta+\nu \right)H_{z}
    &= 0\,
    .
    \end{align}
instead of Eq.~\eqref{4:3}. Its general solution reads
    \begin{align}
    \label{5:5}
    %\label{AirySol}
    H_{z}(\xi) &= C_1\Ai(\nu +\zeta) + C_2\Bi(\nu +\zeta)\,
    .
    \end{align}
Using asymptotic formulas
    \begin{gather}
    \label{5:6}
    \begin{aligned}
    \Ai(\zeta)
    &\approx
    \frac{\e^{-\frac{2}{3}\zeta^{3/2}}}{2\sqrt{\pi}\,\zeta^{1/4}}\,
    \left(
        1
        -
        \frac{5}{48 \zeta^{3/2}}
    \right)
    ,
    \\
    \Bi(\zeta)
    &\approx
    \frac{\e^{\frac{2}{3}\zeta^{3/2}}}{\sqrt{\pi}\,\zeta^{1/4}}\,
    \left(
        1
        +
        \frac{5}{48 \zeta^{3/2}}
    \right)
    ,
%    \\
%    \Ai'(\zeta)
%    &\approx
%    -
%    \frac{\e^{-\frac{2}{3}\zeta^{3/2}}}{2\sqrt{\pi}\,\zeta^{1/4}}\,
%    \left(
%        1
%        +
%        \frac{7}{48 \nu ^{3/2}}
%    \right)
%    ,
    \end{aligned}
    \end{gather}
for large positive values of the argument, $\zeta \gg 1$, leads to the conclusion that radially evanescent solution occurs if $C_{2}=0$. The boundary condition~\eqref{eq:13-1} then reads
    \begin{align}
    \label{5:11}
    \frac{\Ai'(\nu)}
    {\Ai(\nu)}
    +
    i
    %\left(\frac{k_{0}R}{2}\right)^{1/3} \xi
    q
    =
    0
    \,.
    \end{align}
The asymptote \eqref{5:6} for $\Ai(\zeta)$ is valid in the entire range of $\arg(\zeta)$ except for a vicinity of the ray $\arg(\zeta)=\pm\pi$. Taking also the asymptotic formula
    \begin{gather}
    \label{5:6a}
    \Ai'(\zeta)
    \approx
    -
    \frac{\e^{-\frac{2}{3}\zeta^{3/2}}}{2\sqrt{\pi}\,\zeta^{1/4}}\,
    \left(
        1
        +
        \frac{7}{48 \zeta ^{3/2}}
    \right)
    \end{gather}
for the derivative of $\Ai(\zeta)$ in the same range of $\arg(\zeta)$ we reduce Eq.~\eqref{5:11} to the form
    \begin{gather}
    \label{5:12}
    \sqrt{\nu }
    \left(
        1+\frac{1}{4\nu ^{3/2}}
    \right)
    =
    i
    %\left({k_{0}R}/{2}\right)^{1/3} \xi
    q
    \end{gather}
for $|\nu |\gg1$. Solving it same method as we used for Eq.~\eqref{6:14} yields
    \begin{align}
    \label{5:16}
    \Lambda
    =
    -
    \frac{\xi^{2}}{2}
    %-
    +
    \frac{i}{2k_{0}R\xi}\,
    .
    \end{align}
Note that radiative (last) term in \eqref{5:16} has a sign opposite to that in Eq.~\eqref{6:16}. Hence, now this term enhance damping of SW in the direction of propagation. Numerical simulation shows that SW exists for any value of $q$. The effect of curvature in negligible as long as $\left({k_{0}R}/{2}\right)^{1/3}|\xi|\gg1$ and is dominant if $|q|<1$.

Creeping modes are not found in the concave geometry.

\section{Connection to the exact wave equation}
\label{s7}

%$\nu=kR/k_{0}R-1=-\Lambda$, $\rho=k_{0}(R+x)$, $n=kR$

Main results of this paper can be recovered from a solution of the wave equation \eqref{wave_equations'}, which can be solved in terms on the Hankel function of the first kind
    \begin{gather}
    %\label{7:1}
        H_{z}(r) = \Hankel_{n}^{(1)}(\rho)
    \end{gather}
and the second kind
    \begin{gather}
    %\label{7:2}
        H_{z}(r) = \Hankel_{n}^{(2)}(\rho)
        ,
    \end{gather}
respectively for the outgoing and incoming waves; here $\rho=k_{0}r$ and $n=kR$. Corresponding dispersion equations read
    \begin{gather}
    \label{7:3}
      \frac{
        \Hankel_{n}^{(1)\prime}(k_{0}R)
      }{
        \Hankel_{n}^{(1)}(k_{0}R)
      }
      +
      i\xi\,
      =
      0
    \end{gather}
and
    \begin{gather}
    \label{7:4}
      \frac{
        \Hankel_{n}^{(2)\prime}(k_{0}R)
      }{
        \Hankel_{n}^{(2)}(k_{0}R)
      }
      +
      i\xi\,
      =
      0
      ,
    \end{gather}
where the prime stands for the derivative with respect to the function argument. A connection with the content of our paper can de drawn from the asymptotic formulas for the Hankel functions:
    \begin{gather}
    \label{10:7}
    \Hankel_{n}^{(1,2)}(\rho)
    =
    \left( 2/\rho  \right)^{1/3}
    \left[
        %\Ai\left((2/\rho )^{1/3}\tau  \right)
        \Ai\left((2\rho^{2})^{1/3}\tau  \right)
        \mp
        i\,
        %\Bi\left((2/\rho )^{1/3}\tau  \right)
        \Bi\left((2\rho^{2})^{1/3}\tau  \right)
    \right]
    ,
    \end{gather}
where the upper sign corresponds to $\Hankel_{n}^{(1)}(\rho)$, the lower to $\Hankel_{n}^{(2)}(\rho)$, $n=kR=k_{0}R\left(1+\Lambda\right)$, $\tau=n/\rho-1\approx \Lambda - x/R$ and, hence, $(2\rho^{2})^{1/3}\tau \approx (2k_{0}^{2}R^{2})^{1/3}\tau \approx \nu-\zeta$. In our ordering scheme, $\tau \sim \delta^{2}$, $\rho\sim\delta^{-3}$, and $(2\rho^{2})^{1/3}\tau \sim \delta^{0}$. As to validity of Eq.~\eqref{10:7}, its derivation assumes that $\rho\gg1$ and $|\tau|\ll1$. For $\rho\gg|n|$, a different asymptotes should be used, namely
    \begin{gather}
    \label{7:15}
    \Hankel_{n}^{(1,2)}(\rho)
    \sim
    \sqrt{
        \frac{2}{\pi \rho }
    }
    %\exp\left[i\rho  -i\left(2n + 1\right)\pi/4\right]
    \exp\left[\pm i(\rho - \pi n/2 + n^{2}/2\rho - \pi/4)\right]
    %\exp\left[+i n^{2}/2\rho\right]
    .
    \end{gather}
It can readily be derived from the Debye asymptotes of the Hankel functions (see e.g. Appendix of Ref.~\cite{Chen1964JMP_5_820}). Note that the summand $n^{2}/2\rho$ in the exponent is small as compared to $\rho$ if $\rho\gg|n|$, however it can be neglected only if $\rho\gg |n^{2}|$, when it becomes smaller than $1$. Its real part
    \begin{gather*}
    %\label{7:16}
    \re(n^{2}/2\rho)\approx k_{0}R^{2}/2r
    \end{gather*}
gives insignificant phase shift but its imaginary part
    \begin{gather*}
    %\label{7:17}
    \im(n^{2}/2\rho)\approx \im(k)R^{2}/r
    \end{gather*}
is responsible for strong radial variation of the eigen waves.  Because of this summand, the Hankel function $\Hankel_{n}^{(1)}(\rho)$ of complex order with $\im(n)=\im(k)R > 0$ turns out to be fast growing function of its argument in the entire interval
    \begin{gather}
    \label{7:18}
        R\lesssim r \lesssim \im(k)R^{2}
        .
    \end{gather}
In particular,
    \begin{gather*}
    %\label{7:19}
    |\Hankel_{n}^{(1)}(k_{0}r)|
    \sim
    \sqrt{
        \frac{2}{\pi k_{0} R }
    }
    \exp\left[
        \im(k)x
    \right]
    ,
    \end{gather*}
if $x=r-R \ll R$.  All the creeping waves with the eigenvalues \eqref{4:27} are radially rising modes in the interval \eqref{7:18}. They are transformed into algebraically decaying modes with
    \begin{gather*}
    %\label{7:21}
    |\Hankel_{n}^{(1)}(k_{0}r)|
    \sim
    \sqrt{
        \frac{2}{\pi k_{0} r }
    }
    \end{gather*}
at $r\gtrsim \im(k)R^{2}$. Hence, a creeping mode number $j$ is peaked at
    \begin{gather}
    \label{7:22}
    r_{j}
    \sim
    \im(\Lambda_{j})k_{0}R^{2}
    \sim
    (k_{0}R)^{1/3}j^{2/3}R
    .
    \end{gather}
Hence, the creeping waves in the class of Hankel functions $\Hankel_{n}^{(1)}(\rho)$ of the first kind should be ignored when searching for SW localized at the metal-air interface.

%Since $\im(k)=k_{0}\im(\Lambda)>0$ for the the waves under consideration (see Eq.~\eqref{4:34}), it follows from Eq.~\eqref{7:15} that a mode, which is described by the Hankel function $\Hankel_{n}^{(1)}(\rho)$ of the first kind, exponentially grows at the intermediate distances from the metal-air interface and, hence, should be thrown away when searching for a SW localized at the interface. A different reasoning was used in Refs.~\cite{Hasegawa+2004ApplPhysLett_84_1835, Hasegawa2007PhysRevA.75.063816}, where $\Hankel_{n}^{(1)}(\rho)$ was chosen because it describes an outgoing wave at $r\to\infty$. We have shown, however, that an outgoing wave is formed at extremely large distances $r\gg (k_{0}R)^{1/3}R$ from the metal-air interface.

As to the wave $\Hankel_{n}^{(2)}(\rho)$ of the second kind, it exponentially decreases in the interval of radii \eqref{7:18} provided that $\im(k)>0$. Hence, it is suitable for a SW localized at the metal-air interface. As we can deduce from our study of Eq.~\eqref{4:11e}, its full-scale analog \eqref{7:4} can have one and only one root with $\im(k)>0$. It is this root, which stands for SW. At large distances from the interface, $r>r_{\Lambda}=k_{0}\im(\Lambda)R^{2}$ with $\Lambda $ given by Eq.~\eqref{6:16}, this wave is transformed into freely propagating incoming cylindrical wave. For this reason a solution containing $\Hankel_{n}^{(2)}(\rho)$ was thrown away in Refs.~\cite{Hasegawa+2004ApplPhysLett_84_1835, Hasegawa2007PhysRevA.75.063816, Liaw+2008OE.16.4945, Guasoni2011JOSAB.28.1396, Guo+2009IEEE.8.408}. We note however that this reason seemed not convincing in case if $k_{0}\im(\Lambda)R\gg1$ and the transformation occurs at exponentially low amplitude of the wave, $|\Hankel_{n}^{(2)}(k_{0}r_{\Lambda})/\Hankel_{n}^{(2)}(k_{0}R)| \sim \exp(-k_{0}\im(\Lambda )R)$.

\section{Summary}
\label{s8}

In the majority of the literature on the subject of electromagnetic waves propagation in the vicinity of a conducting cylinder, the Hankel functions of the second kind are discarded with the argumentation that they describe waves that asymptotically propagate toward the cylinder. One of the main results of our paper is that we show that for a boundary condition corresponding to a lossy boundary, this is, in general, incorrect. Indeed, the Hankel functions in this case have large complex indices, and their asymptotic behavior is more complex than just ``outgoing'' or ``incoming'' wave criterion. We consider the distribution of the Poynting vector in the vicinity of the cylinder, and make our conclusion on localization of a particular mode with account of the energy flow in the mode.

Our approach uses a paraxial equation \eqref{master_equations} to describe a surface wave at a cylindrical metal-air interface in the low frequency limit with $\omega \ll\omega_{p}$. Its solution is found in terms of the Airy-Fock functions \eqref{4:42} of the first and second kind, avoiding in this way the complexities of dealing with the Hankel functions of large argument and large order used in earlier studies. We have shown that surface waves of two types can exist on a curved metal boundary which we called the outgoing and incoming surface waves. Inspecting the radial component $S_{r}$ of the Poynting vector, we reveal a qualitative difference between the two types of the surface waves. In the vicinity of the metal-air interface, the radial energy flux is directed inwards for the waves of both types as expected for a lossy medium (ie, a realistic metal). However $S_{r}$ reverses its sign at large distances from the interface in case of the outgoing surface wave.
 We also emphasized that only incoming surface wave survives in the limit of planar metal-air boundary---this fact, being always known, was not properly evaluated by those authors who discarded incoming solutions in the cylindrical geometry.

We have derived the dispersion relations for the waves of both types by computing the radiative corrections to the dispersion relation of a planar surface wave due to the curvature of the metal-air interface, see Eqs.~\eqref{4:24} and~\eqref{6:16}.
%We suppose that the surface wave of the second kind was overlooked in the earlier studies \cite{Hasegawa+2004ApplPhysLett_84_1835, Hasegawa2007PhysRevA.75.063816, Liaw+2008OE.16.4945, Guasoni2011JOSAB.28.1396, Guo+2009IEEE.8.408} although is it the only type that survives in the planar geometry.
We found that for any value of the surface impedance there is at most one surface mode of both types. We believe that some of the modes previously reported as the surface waves are indeed the so called creeping waves. %The creeping modes have radially growing amplitudes in the vicinity of the metal-air interface; they assume maximal amplitude at large distances from the interface.

\section*{Acknowledgements}
%\begin{acknowledgments}
    %The work was supported by the Department of Energy, contract DE-AC03-76SF00515, and by the Ministry of Education and Science of Russian Federation, project RFMEFI61914X0003.

    The authors are grateful to B.A. Knyazev, I.V. Andronov and fourth Referee for useful references.
%\end{acknowledgments}

%%\bibliographystyle{osajnl}
%%%\bibliographystyle{apsrev4-1} % makes an error for me
%%%\bibliographystyle{apsrmp4-1}
%\bibliography{../SEW}

%\bibliography{../../SEW}

\end{document}